# Dose finding for new vaccines: the role for immunostimulation/immunodynamic modelling

Authors: Sophie J. Rhodes[a]*, Gwenan M. Knight[a], Denise E. Kirschner[b], Richard G. White[a+], Thomas G. Evans[c+]

* Corresponding author

+ joint senior authors

Institutions:

[a] TB Modelling Group, CMMID, TB Centre, London School of Hygiene and Tropical Medicine, UK

[b] University of Michigan Medical School, Ann Arbor, MI, USA

[c] Vaccitech, Oxford, UK

**Abstract**

Current methods to optimize vaccine dose are purely empirically based, whereas in the drug development field, dosing determinations use far more advanced quantitative methodology to accelerate decision-making. Applying these established methods in the field of vaccine development may reduce the currently large clinical trial sample sizes, long time frames, high costs, and ultimately have a better potential to save lives. We propose the field of immunostimulation/immunodynamic (IS/ID) modelling, which aims to translate mathematical frameworks used for drug dosing towards optimizing vaccine dose decision-making. Analogous to PK/PD modelling, IS/ID modelling approaches apply mathematical models to describe the underlying mechanisms by which the immune response is stimulated by vaccination (IS) and the resulting measured immune response dynamics (ID). To move IS/ID modelling forward, existing datasets and further data on vaccine allometry and dose-dependent dynamics need to be generated and collate, requiring a collaborative environment with input from academia, industry, regulators, governmental and non-governmental agencies to share modelling expertise, and connect modellers to vaccine data.

**Importance of vaccine dose-response curve shape**

Vaccines are a key public health discovery and are one of the most cost-efficient interventions available in medicine [1]. Finding optimal vaccine dose amounts (hereafter dose), as well as identifying appropriate regimens, are key factors in reaching maximal vaccine efficacy at the requisite safety level. However, taking a vaccine from discovery to licensure can cost in the range of US$0.8 billion [2]. With these enormous costs, there is an intense pressure to make well-informed decisions at each stage of the development process. However, the current use of antiquated methods, may lead to sub-optimal dosing decisions.

Recent conditions including increased demand, resource limitation and cost have led the World Health Organization (WHO) to conduct retrospective dose ranging studies where the immunogenicity and/or efficacy of fractional doses (of full licensed doses) were tested. For diseases such as yellow fever [3, 4], meningitis [5] and malaria [6], assuming the same administration route, fractional doses were found to be equivalent to, or in some cases more immunogenic than full licensed doses. Retrospective dose ranging studies have led to a policy change in yellow fever vaccines, where the WHO now recommends that a fractional 1/5[th] (volume) yellow fever vaccine dose may be used in outbreak situations where supply is low [7]. This fractional dose has been administered to five million persons in Brazil to date [8]. These studies highlight an important question; if smaller doses are optimal with respect to immunogenicity (and participant exposure to pathogen) as compared to a large licensed dose, then why were they either missed or not initially selected for licensure during development?

Another example of sub-optimal dosing decisions is evident in the development of the novel subunit TB vaccines, H-series, which are currently in phase 1/2a clinical trials. Development of the H-series vaccine has benefitted from robust dose escalation studies. The dose ranging data from small animal studies suggested that the immunogenicity of the vaccine was highest at middle doses (0.05 to 1 µg vaccine antigen) and then decreased with the higher doses (5 and 15 µg vaccine antigen) [9-11]. This phenomenon creates a peaked or n-shaped dose-response curve, commonly referred to as the goldilocks effect [12]. Clinical testing showed a similar peaked dose-response curve; smaller doses (5, 15 µg vaccine antigen) were more immunogenic than higher doses (50, 150 µg) [13, 14]. However, in the early phase 1/2 clinical trials, a single dose of 50 μg vaccine antigen was chosen [15-17]. Thus, despite the pre-clinical dose-response data showing lower doses were more immunogenic, the lower end of the dose-response curve in humans has not been fully explored and higher doses were selected.

Peaked dose-response curves are not unique to TB vaccines; similar dose-response curves have also been seen in adenovirus (Ad35) [18, 19], HIV [20], malaria [6] and influenza [21] vaccines. This is contrary to a long-standing vaccine development assumption that the relationship between dose and host response is saturating (sigmoidal) [22]; i.e. there is a minimum vaccine dose that gives no host response, a window of vaccine doses where the response rapidly escalates and a plateau above a certain dose threshold. Following this assumption, the goal of vaccine development is to then increase dose until a response plateaus and assume that it is the highest, safe optimal dose (with some margin of error to allow for host variation). In contrast, peaked curves suggest that there is a risk that high, suboptimal doses could be progressing to later clinical development stages. Thus, more in-depth analysis of the shape of the vaccine dose-immune response curve, which has been largely ignored to date, is essential to understanding how to select optimal dose.

**Difficulties with current vaccine dose-finding approaches**

It is likely that in many cases, sub-optimal vaccines doses have been selected. Surprisingly, the definitive text on vaccine development does not include strategies for dose finding [23] and there is limited regulatory guidance on dose-finding methodologies from licensing organizations such as the FDA.

Currently, estimates for effective human doses are based on responses in small animal models (such as mice and rats) in which large dose ranges are tested over short timeframes. Typically, "low" doses that are used in mice or other small animals are selected and increased by half log to one log increments until an assumed maximum plateau in response is met. The next step is then to translate vaccine responses from these animal studies to humans, known as allometric scaling. Briefly, allometric scaling is the quantifiable relationship between animal body size and characteristic, e.g. the physiological relationship between animal size and metabolism or life span. In humans, allometric scaling is applied to common PK parameters such as volume of distribution, absorption and clearance by using the host's weight, (e.g. for the drug Isoniazid e.g.[24]). Challenges are faced when applying vaccine dose allometric scaling between species, as the immunological relationships are still not well characterized, and fraught with issues of not only scale, but physiological differences between species. For example, assumed vaccine dose allometric scaling factors between mice and human vary over large ranges from 5-20 for HPV vaccines [25, 26] to 0.5-100 for TB vaccines [27-30]. There is a significant gap and lack of research into how vaccine doses translate across species and current vaccine dose allometric scaling assumptions are commonly not explicitly disclosed.

Additional technical issue are present with vaccine development studies. First, an inability to dilute vaccines to small enough dose, lack of more than one dose formulation starting material, and/or assay variation could also be contributing to under-researched dose-response curves. For example, a trial of a gp120 vaccine for HIV-1 infection in humans with different adjuvants, revealed that the surrogate response (which in this case was binding and homologous virus neutralization) at a dose of 30 μg formulated with QS-21 was equivalent to that of 300 μg of the same vaccine in alum. A further study with 0.5, 3, and 30 μg of the vaccine in the adjuvant, QS-21, revealed no decrease in response. Thus, neither the lower bound nor the shape of the dose-response curve has been established, primarily due to an inability to further accurately dilute this vaccine. On the other end of the dose-response curve, accurate assessment can be limited by an inability to achieve sufficient concentrations to reach a maximal dose as defined by immunologic, clinical, or safety parameters.

Second, a common barrier to efficient vaccine development is the potential lack of biological marker of protection, or biomarker. Without such a biomarker, validation of early decisions on developmental variables, such as dose, are unattainable. However, vaccines are often progressed through to clinical trials based on a hypothetical surrogate of protection - only after efficacy has been confirmed can a chosen surrogate be properly validated [31].

In summary, it is likely that the current empirical methods used in vaccine dose finding are leading to sub-optimal vaccines dose selections. Thus, to identify an optimal combination of developmental variables with current approaches (e.g. dose, dose regimen, vaccine composition (adjuvant dose)) will require a large, expensive multi-dimensional factorial design trial. As a result, vaccine doses are moving forward without extensive evaluation, which is often due to insufficient funding, developmental time pressures and lack of a clear optimized pipeline. Can a more effective and systematic identification protocol for optimal vaccine dosing be achieved?

Immunostimulation (IS) /Immunodynamic (ID) modelling: Mathematical modelling for improved vaccine dose decision making

The world of drug development has faced similar drug-dosing questions, yet is far more advanced in the use of systematic methods for dose optimization. This can be attributed partly to the use of pharmacometrics (or systems pharmacology): mathematical models that describe within host drug dynamics. The most commonly models used are pharmacokinetic/pharmacodynamic (PK/PD) models that employ mechanistic mathematical models to quantify drug concentration dynamics within the host over time (PK) and track drug effect and dynamics as drug concentration varies (PD) [32]. Model-Based Drug Development (MBDD) is recognized as an efficient tool to accelerate and streamline drug development by minimizing developmental time and resources and is regularly used to establish optimal drug doses [33]. MBDD has been established for decades in the pharmaceutical industry to improve dose selection for small molecule drugs [34] and is often required by regulatory agencies in all stages of drug development. As an example, modelling was able to tease through the different doses and protocols to derive optimal values for TB drug treatments, which previously had never been formally compared [35].

As yet, there is no such parallel to these methods used for vaccine dosing, which may be due to the diversity and complexity of immune responses measured or a lack of appreciation of potential quantitative tools. Application of similar methods to those used in drug development could lead to better evaluation of vaccine dose-response data derived from animals and its translation to humans to potentially improve vaccine development. Vaccine development is now in a position (decades later) to borrow from the experiences, expertise and technical utilities of MBDD. Consequently, we propose the new field of vaccine *Immunostimulation/Immunodynamic (IS/ID) modelling* as a method to improve vaccine dose decision-making and ultimately vaccine discovery. Analogous to PK/PD modelling, IS/ID modelling applies mathematical models to describe the underlying mechanisms, the immune response stimulation (IS) that produces measured immune response dynamics following vaccination (ID).

The application of IS/ID modelling to accelerate vaccine development could provide the following benefits. Firstly, a mathematical and computational modelling framework would have the capability to more effectively evaluate *in silico* greater combinations of vaccine trial design variables and narrow the design space before trials ever begin. Secondly, the incorporation of mathematical modelling into pre-clinical vaccine development could eventually result in a reduction in laboratory animals by replacing empirical experimentation with *in silico* simulation that optimizes the selection of doses and number of animals [36]. Similarly, the application of IS/ID modelling to clinical vaccine trial design could also reduce trial sample size and thus the total human exposure to investigational agents. Finally, the immune response required for protection against a disease relies on complex interactions that behave nonlinearly over time and across multiple biological scales (e.g. molecular to cellular to whole systems). IS/ID models will allow us to quantify this complexity to obtain meaningful biological predictions while retaining model identifiability.

IS/ID modelling vaccine development is limited only by the issues already facing the vaccine development world. Decisions on vaccine dose and dose regimen are currently being made regardless of a developer's knowledge on dose allometry, an established correlate of protection, or extensive immune response data. Our goal with IS/ID modelling is not to discuss the correct vaccine induced immune measure, but to suggest a more *systematic framework* to inform key vaccine dose decisions based on a chosen surrogate and applying established methods currently used in drug development. Consequently, we would like to highlight that application of IS/ID modelling to vaccine development is not limited to a particular disease, type of vaccine or induction of a specific type of immune response and is also adaptable to vaccine requirements and available data.

IS/ID modelling implementation

The generalized steps (for any vaccine) to integrate modelling into dose finding are outlined below, (a scheme of the steps is in Figure 1):

1. A wide range of doses of a new immunogenic vaccine are tested in small animal models to find minimum and maximum doses that provide the bounds of the dose-response curve (note adjuvant dose could/should also be varied). The dose range, although wide, can be based on historical work of similar vaccines.
2. IS/ID mathematical modelling is applied to estimate the parameters that describe the underlying dynamics of the initial animal-derived dose-response relationship. Optimal experimental design is then generated to yield the maximum information on the dose-response curve (with a pre-specified confidence interval), given limitations on animal numbers, ability to achieve desired concentrations of the product, and cost.
3. The IS/ID model is calibrated to human response data on limited doses and using allometric dose scaling assumptions (or tested in a human-immune response computational model [37]), the animal IS/ID model parameters are used to predict the theoretical human dose-response relationship and can be tested in the human-immune response computational model.
4. As in step 1, a selection of doses are chosen to define an approximate shape and the confidence bounds of the human dose-response curve based on the theoretical prediction in steps 1-3.
5. These data are then fed back into the model to gain understanding of the confidence intervals around the chosen doses. As further human data are collected, the IS/ID model is refined and used to hone in on best dose and it's confidence interval.

**Current IS/ID modelling**

The first steps to include IS/ID modelling into vaccine development are under way. We conducted an intensive animal vaccine multi-dose study of a candidate TB vaccine (presently in phase 2) designed by Rhodes et. al. to specifically generate data for dose-response identification using translational IS/ID methods. IS/ID modelling was applied to determine the dose-response curve and showed a definitive n-shape for multiple times points after vaccination [11]. This strongly suggests that the predicted best dose is likely to be lower than previously investigated [11]. Preliminary work has also been conducted to simulate further responses (based on these empirical data) to obtain optimal dosing cohorts and narrow the confidence interval around predicted best dose. In addition, current work underway develops an IS/ID mechanistic model of the TB vaccine dose-dependent induction of memory, effector, and regulatory T cell subsets and, using well established model parameter estimation methods employed in population PK/PD modelling (i.e. nonlinear mixed effects modelling), translated the dose-response curve from mice to humans ([38] and paper in submission). The results of this work suggest a lower dose than has been used in phase 1 trials in humans may be the most immunogenic. Importantly, this model prediction has since been supported by large phase 2 trial.

## Future work to inform IS/ID vaccine dose modelling

In order to apply IS/ID modelling effectively to vaccine development, further datasets are required. For example, a thorough investigation into vaccine allometry, which is vital to scale vaccine dose across species, should be undertaken. Additionally, more extensive studies on immune system dynamics by dose should be introduced early on to pre-clinical investigations. In drug development, modelers use all available (relevant and standardized) data to refine understanding of PK/PD findings throughout product development. To maximize information on vaccine allometric scaling information and model parameterization across vaccines in the same way, we recommend the creation and curation of large shared dose-response data platforms through large-scale collaborative efforts with multinational pharmaceutical companies.

Second, a collaborative group of interested parties from academia, biotech, large vaccine manufacturers, regulators, governmental and non-governmental agencies must be established to aid communication, data access and development of methodology. The first meeting of such a group of individuals occurred in May 2015 at the headquarters of TB vaccine developers Aeras (Rockville, MD), where a multidisciplinary team met to discuss the current state of vaccine dose finding and the potential for mathematical modelling to assist in this arena. Creating such a network, would provide links between existing modelling consortia such as the TB, HIV, Malaria, NCD modelling consortia and International Society of

Pharmacometrics to facilitate access to modelling expertise. Incentives such as open access to large historical data packages, and a commitment by vaccine developers, should be put in place to encourage modelers with experience in drug dosing and immune response modeling to move to vaccine development, which can have a potentially greater impact on human health. We believe that vaccine regulatory bodies also lack critical evaluations of vaccine product dose selection, and that agencies such as FDA or EMA should encourage modelers to move from the drug development focus into vaccine development. Thus, the motivation and investments must come in both a bottom-up and top-down fashion.

Finally, and most critically, we encourage the NIH and other vaccine funding agencies to consider head to head studies in which conventional methods for selecting vaccine do are used in parallel with the outlined modelling techniques, to better understand the impact on speed of development, number of participants exposed and cost of vaccine design. Only then can the true value of the endeavor be assessed.

## Funding

SR is funded by a studentship from Aeras. RGW is funded the UK Medical Research Council (MRC) and the UK Department for International Development (DFID) under the MRC/DFID Concordat agreement that is also part of the EDCTP2 programme supported by the European Union (MR/P002404/1), the Bill and Melinda Gates Foundation (TB Modelling and Analysis Consortium: OPP1084276/OPP1135288, CORTIS: OPP1137034, Vaccines: OPP1160830) and UNITAID (4214-LSHTM-Sept15; PO 8477-0-600).

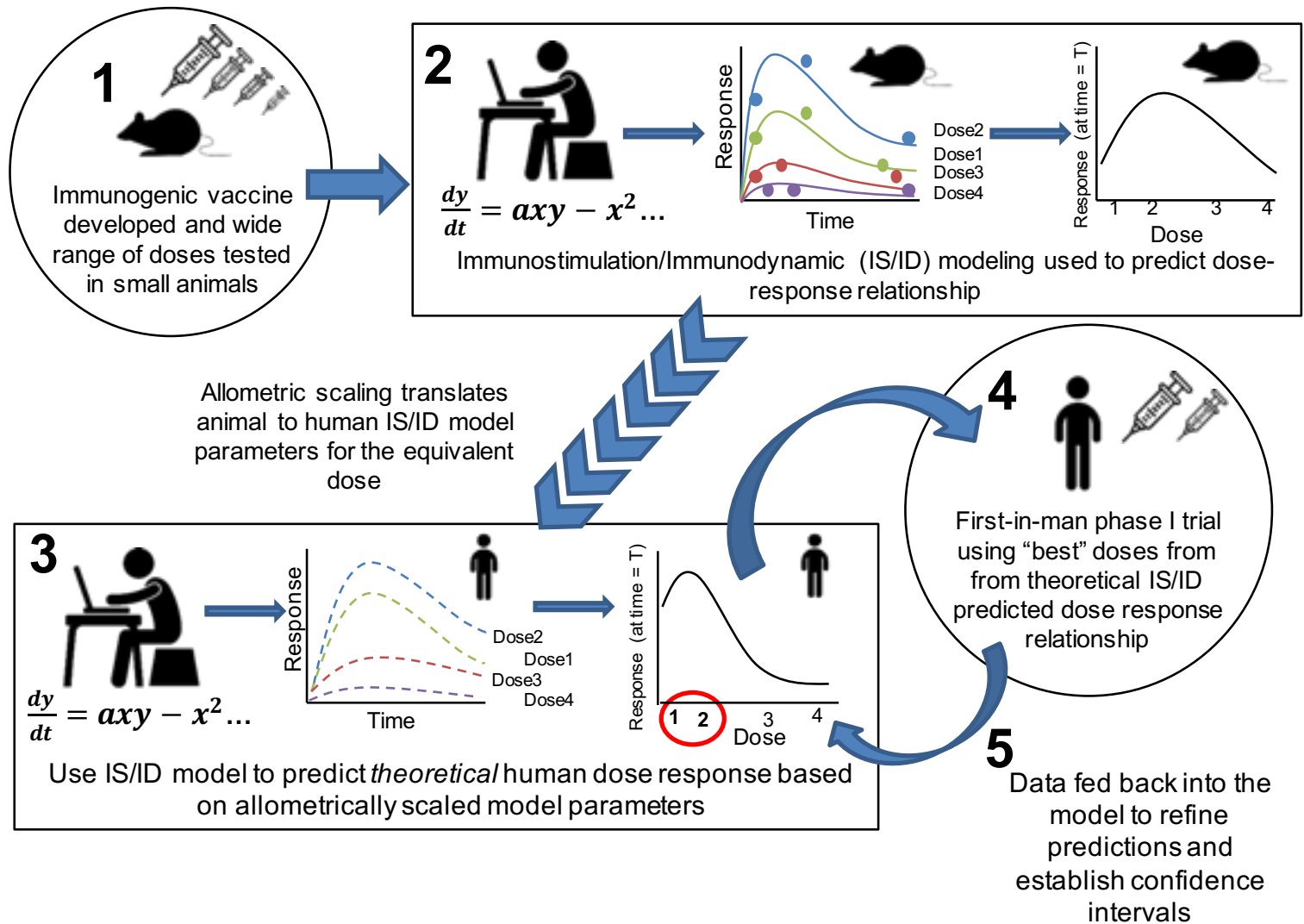

Figure 1. Schema depicting the steps required to incorporate immunostimulation (IS) /immunodynamic (ID) modeling into vaccine development. [Graphics included credited to: https://thenounproject.com].